\documentclass[preprint,aps,amsmath,amssymb,prd,floatfix]{revtex4-1}
\usepackage{graphicx}
\usepackage{hyperref}
\DeclareMathOperator{\IM}{Im}
\begin{document}

\title{Holographic Non-Fermi Liquids and the Luttinger Theorem}
\author{Finn Larsen and Greg van Anders}
\affiliation
{Michigan Center for Theoretical Physics, Randall Laboratory of Physics,\\
The University of Michigan, Ann Arbor, MI 48109-1040, USA}
\begin{abstract}
We show that the Luttinger theorem, a robust feature of Fermi liquids, can be
violated in non-Fermi liquids. We compute non-Fermi liquid Green functions using
duality to black holes and find that the volume of the Fermi surface depends 
exponentially on the scaling dimension, which is a measure of the coupling. 
This demonstrates that Luttinger's theorem does not extend to non-Fermi liquids.
We comment on possible experimental signatures.
\end{abstract}
\maketitle
%\paragraph*{Introduction} \label{intro}
\section{Introduction} \label{intro}
Though many metals are well described by Landau Fermi-liquid theory, there are
numerous examples of materials that are not well-described by weakly interacting
quasiparticles \cite{nFLrev}. Non-Fermi liquids, which include the normal phase
of of high-$T_c$ superconductors, are inherently strongly coupled, which makes
them interesting systems to study. Luttinger's theorem \cite{luttthm}, the
constancy of the phase-space volume contained within the Fermi surface as a
function of coupling, is a robust feature of Fermi liquids. It is not
{\it a priori} obvious, however, that this theorem persists in non-Fermi
liquid systems, even when a notion of a Fermi surface does apply. In this
Letter we show that Luttinger's 
theorem is violated in a particular non-Fermi liquid system.

Since non-Fermi liquids are inherently strongly coupled, finding tractable
fora for investigating them is difficult. Beginning with \cite{sslee} there
has been considerable interest in studying strongly coupled fermions using
gauge/gravity duality \cite{adscft1}. Although there have been suggestions of the
formation of Fermi surfaces in string-theoretic brane constructions
\cite{baryon,sv}, the bottom up methods for studying probe fermions in
charged black hole backgrounds developed in \cite{sslee,lmv,czs,flmv} 
provide the most convenient setting for studying non-Fermi liquid behavior.

In this Letter we use holographic techniques to show that Luttinger's theorem
can be violated in non-Fermi liquids. We study fermionic correlation functions
in anti-de Sitter- (AdS) black hole backgrounds using probe fermions
of various masses. Via the holographic dictionary for fermions \cite{mueckvish} the conformal
dimension of the dual operator in the boundary conformal field theory (CFT) is
controlled by the mass of the bulk fermion. In \cite{czs} it was found that, if
the mass was tuned so that the conformal dimension of the operator in the
boundary theory coincided with that of a free fermion, the spectral function
exhibited a peak consistent with that of a Fermi liquid. Tuning the mass so that
the conformal dimension departed from the free value yielded behavior that
deviated from the Fermi liquid. The authors of \cite{czs} interpreted the mass,
then, as a proxy for the coupling, and found that the Fermi momentum remained
constant as the coupling was varied, as predicted by Luttinger's theorem.

However, the fermionic spectral function in the background of a charged black
hole has multiple peaks, and the peak under consideration in \cite{czs} is
different than the non-Fermi liquid peak considered in \cite{lmv}. In this
Letter we study the spectral peak in \cite{lmv} as a function of the fermion
mass. Following \cite{czs}, we interpret the mass as a proxy for the coupling.
We find that the Fermi momentum depends exponentially on the probe fermion
mass, and thereby on the coupling, in clear violation of Luttinger's theorem.

We consider boundary CFTs that are both $2+1$- and $3+1$-dimensional. In
the former case, the result persists in the presence a magnetic
field \cite{HKdyonic}. In that case, the bulk wave equation can be expanded in
terms of Landau levels, and the resulting wave equation can be reduced to the
electrically charged case \cite{bhms2,dhs}. As we will discuss, this allows for
a possible experimental signature of this effect in de Haas-van Alphen
measurements.

%\paragraph*{Setup}\label{setup}
\section{Setup}\label{setup}
We start with the Reissner-Nordstr\"om-AdS black hole in $d+1$ dimensions with
the metric
\begin{equation}\label{metric}
  ds^2 = -f(r)dt^2+\frac{dr^2}{f(r)}+\frac{r^2}{L^2}dx_i^2 \, ,
\end{equation}
with (e.g.\ \cite{cejm})
\begin{equation} \label{fA}
  \begin{split}
    f(r)&=\frac{r^2}{L^2}-\frac{M}{r^{d-2}}+\frac{Q^2}{r^{2(d-2)}}\, ,\\
    A&=\left(\mu-\sqrt{\frac{d-1}{2(d-2)}}\frac{Q}{r^{d-2}}\right)dt\, .
  \end{split}
\end{equation}
We introduce an orthonormal frame according to
\begin{equation}\label{frame}
  e^0=\sqrt{f(r)}dt\, , \qquad e^{i}=\frac{r}Ldx^i\, , \qquad
  e^d=\frac{dr}{\sqrt{f(r)}} \, ,
\end{equation}
where $i=1,\dotsc,(d-1)$. The spin connection is 
\begin{equation}\label{spinconnection}
  \omega^{ab}= \frac{f'(r)}2(\delta^a_0\delta^b_d-\delta^a_d\delta^b_0)dt
  +\frac1L\sqrt{f(r)}(\delta^a_i\delta^b_d-\delta^a_d\delta^b_i)dx_i \, .
\end{equation}
We will scale $Q$, $M$ by a dimensionful parameter $r_h$ as 
\begin{equation}\label{rescale}
  Q\to \frac{r_h^{d-1}}LQ \, ,\qquad
  M\to \frac{r_h^d}{L^2} M \, .
\end{equation}
If the dimensionless charges are related by $M=1+Q^2$ then $r_h$ is the largest
zero of $f(r)$ and therefore the horizon radius. We also adjust $\mu$ so that the
gauge field vanishes at the horizon
\begin{equation}
  A=\mu\left(1-\frac{r_h^{d-2}}{r^{d-2}}\right)dt\, , \qquad
  \mu = \sqrt{\frac{d-1}{2(d-2)}}Q \frac{r_h}L \, .
\end{equation}
The temperature of the black hole is
\begin{equation}
  T=\frac{r_h}{4\pi L^2}(d-(d-2)Q^2) \, ,
\end{equation}
whereby the extremal black hole has $Q^2 = \frac{d}{d-2}$.

The action for a minimally coupled fermion is
\begin{equation}\label{diracaction}
  S_\text{Dirac}=\int d^{d+1}x\sqrt{-g}i(\bar\psi\Gamma^ae_a^\mu
  D_\mu\psi-m\bar\psi\psi) \, ,
\end{equation}
where
\begin{equation}\label{Ddef}
  \bar\psi=\psi^\dagger \Gamma^0 \quad
  D_\mu=\partial_\mu+\frac18\omega_\mu^{ab}[\Gamma_a,\Gamma_b]-iqA_\mu \, .
\end{equation}
It is convenient to work in terms of eigenspinors $\psi_\pm=\Gamma_\pm\psi$ of
the projection operators $\Gamma_\pm=\frac12(1\pm\Gamma^d)$. We will also
consider $m>0$ without loss of generality.

Defining now
\begin{equation}\label{psiproj}
  \psi_\pm=(f(r))^{-\frac14} r^{-\frac{d-1}2} e^{-i\omega t+i k_i x^i}\phi_\pm
  \, ,
\end{equation}
the Dirac equation has the form
\begin{equation}\label{diraceqn}
  r\sqrt{f(r)}\left(\partial_r\mp \frac{m}{\sqrt{f(r)}}\right)\phi_\pm=
  \mp i\gamma^\mu K_\mu L\phi_\mp \, ,
\end{equation}
where
\begin{equation}\label{Kdef}
  K^\mu=(u,k_i)\, , \quad
  u=\frac{r}{L\sqrt{f(r)}}(\omega+qA_t) \, ,
\end{equation}
and $\gamma^\mu$ are $d$-spacetime-dimensional gamma matrices.

If $mL<\tfrac12$, then near the boundary
\begin{equation}\label{asympt}
  \begin{split}
  \phi_+&\approx
  A\left(\frac r{r_h}\right)^{mL}+B\left(\frac r{r_h}\right)^{-mL-1}, \\
  \phi_-&\approx
  D\left(\frac r{r_h}\right)^{-mL}+C\left(\frac r{r_h}\right)^{mL-1},
  \end{split}
\end{equation}
where the coefficients are related by
\begin{equation}\label{coeffs}
  C=\frac{iL^2\gamma^\mu k_\mu}{(2mL-1)r_h}A,\quad
  B=\frac{iL^2\gamma^\mu k_\mu}{(2mL+1)r_h}D,
\end{equation}
and $k^\mu=(\omega+q\mu,k_i)$. 

Since we have rotational symmetry on the boundary, pick
$\vec{k}=(k,0,\dotsc,0)$, and the gamma matrices so that $\gamma^0=i\sigma_2$,
$\gamma^1=\sigma_1$ and $\phi_\pm=\begin{pmatrix}y_\pm\\z_\pm\end{pmatrix}$. The
Dirac equation can then be written as
\begin{equation}\label{projected}
  \begin{split}
    r\sqrt{f(r)}\left(\partial_r\mp\frac{m}{\sqrt{f(r)}}\right)y_\pm&=
    \mp iL\left(k-u\right)z_\mp\, ,\\
    r\sqrt{f(r)}\left(\partial_r\mp\frac{m}{\sqrt{f(r)}}\right)z_\pm&=
    \mp iL\left(k+u\right)y_\mp\, .
  \end{split}
\end{equation}

Let us henceforth consider the extremal black hole. The leading order equation
of motion for $y_+$ in the near horizon region $r=r_h+L\epsilon$ for
$\epsilon \ll r_h/L$ can be written as
\begin{equation}
  \left(\epsilon^2\partial_\epsilon+\tilde m\epsilon\right)
  \left(\epsilon^2\partial_\epsilon-\tilde m\epsilon\right)y_+=
  -\tilde\omega^2y_+ \, ,
\end{equation}
where $\tilde m=mL/\sqrt{d(d-1)}$ and $\tilde\omega=\omega L/d(d-1)$.
The corresponding equation for $z_+$ is identical. This equation can be solved
exactly in terms of Hankel functions, however we need only the leading small
$\epsilon$ behavior. Selecting the solution that is purely ingoing at the
horizon, appropriate for computing retarded correlation functions, we can choose
overall normalizations so that
\begin{equation}
  y_+\approx z_+\approx y_-\approx -z_-\approx
  e^{i\frac{\tilde\omega}\epsilon} \, .
\end{equation}

The response function (for $mL>0$) is computed as
\begin{equation}\label{Gdef}
  G_R=\lim_{\epsilon\to0}\epsilon^{-2mL}
  \left.\begin{pmatrix}\xi_+&0\\
    0&\xi_-\end{pmatrix}\right|_{r=\frac{r_h}\epsilon} \, ,
\end{equation}
where
\begin{equation}\label{xipdef}
 \xi_\pm = \left(i\frac{y_\mp}{z_\pm}\right)^{\pm1} \, ,
\end{equation}
satisfies the equation of motion
\begin{equation}\label{pnlin}
  r\sqrt{f(r)}\partial_r\xi_\pm=-2r m\xi_\pm \mp L(k\mp u)\pm L(k\pm u)\xi_\pm^2
  \, .
\end{equation}
Ingoing boundary conditions at the horizon take the form
\begin{equation}\label{bc}
  \left.\xi_\pm\right|_{r=r_h}=i \, .
\end{equation}

We will compute the Green function \eqref{Gdef} by solving the equation of
motion \eqref{pnlin} numerically. The ratio \eqref{xipdef} is convenient for this
because, as a ratio of waves oscillating in phase, it is not itself oscillatory. Also, 
since $f(r_h)=0$, for numerical purposes we expand the equation of 
motion near $r=r_h$ and begin the numerical
evolution just outside the horizon.

%\paragraph*{Luttinger Theorem}
\section{Luttinger Theorem}
The Luttinger theorem \cite{luttthm} states that the volume of phase space
contained within the Fermi surface remains constant as the coupling between
quasiparticles is adjusted. In \cite{luttthm}, the Fermi surface was defined as
the location of a discontinuity in the density of states. Standard Landau Fermi
liquid theory \cite{luttthm,llv9} associates this discontinuity with a peak in
the fermion Green function. In isotropic situations, such as the one we are
considering, the Fermi surface is a sphere, which means that constancy of the
volume enclosed by the Fermi surface implies constancy of the Fermi momentum.
If holographic non-Fermi liquids obey the Luttinger theorem, the non-Fermi
liquid peak would remain at a fixed momentum as the coupling is adjusted. We
will see that this is not the case.

Holography associates the scaling dimension of CFT operators with the
asymptotic scaling behavior of the bulk gravity field. Generically, this
scaling behavior is determined by the dimensionality of the bulk, the spin of
the bulk field, and its mass. Since the scaling dimensions of operators flow
with the coupling strength, adjusting the mass of the bulk field, and thereby
its scaling dimension, can be interpreted as adjusting the coupling in the CFT.

Let us concentrate for now on the situation in $AdS_4$-Reissner-Nordstr\"om.
The Green function in this background has been found to exhibit two peaks. One
is the peak in $\IM(G_{--})$ that was identified in \cite{lmv} and is associated
with non-Fermi liquid behavior. The other found in \cite{czs}, is in
$\IM(G_{++})$ \cite{schalm} in this notation, and is associated
with Fermi-liquid behavior. The Fermi-liquid peak studied in \cite{czs} was
found to remain at approximately fixed momentum as a function of the bulk mass,
in agreement with Luttinger's theorem. We find that this does not apply for the
non-Fermi liquid peak.

\begin{figure}
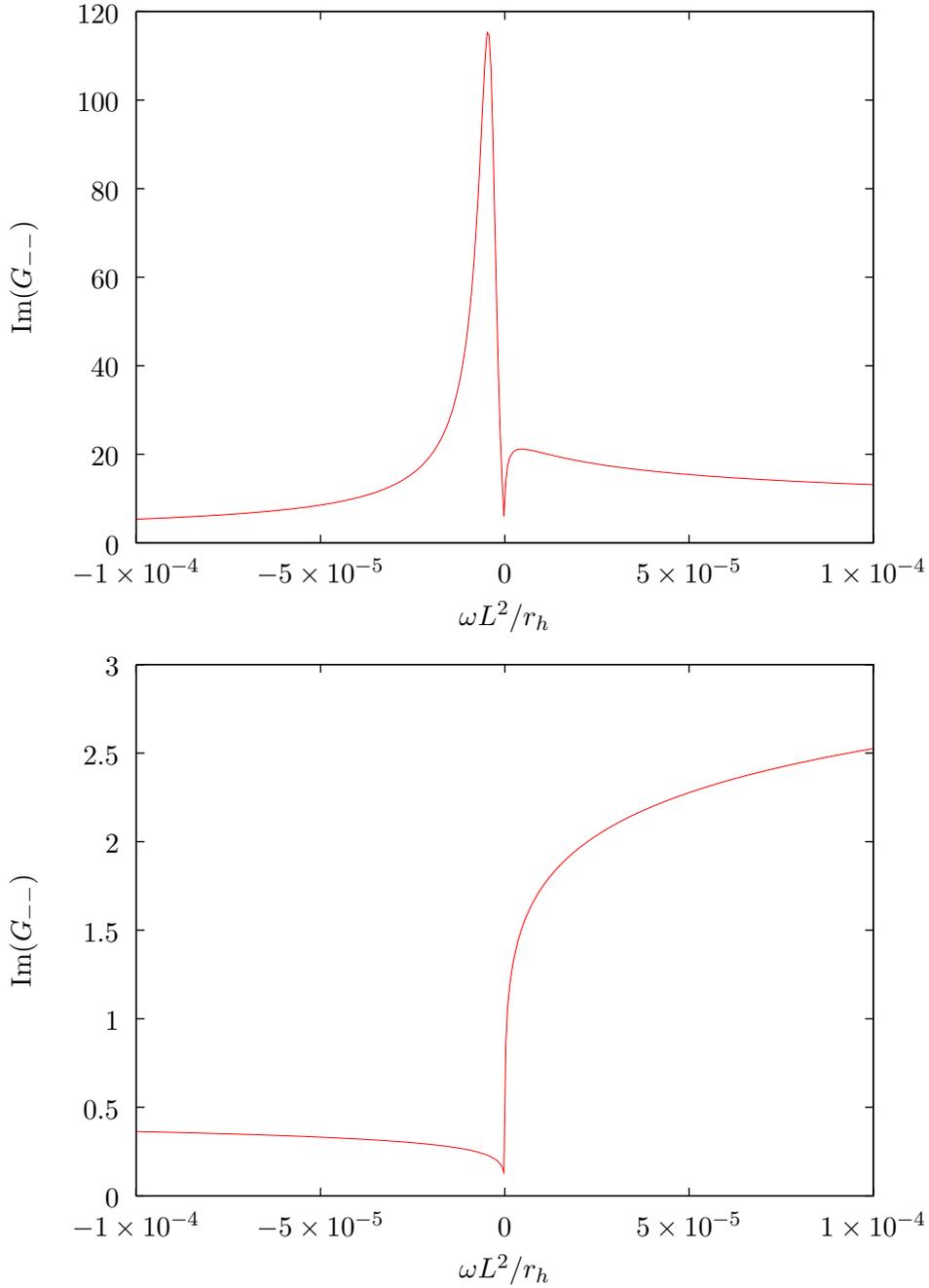

  %\scalebox{0.6}{\input{kltkf.tex}}
  %\scalebox{0.6}{\input{kgtkf.tex}}
  \input{kltkf.tex}
  \input{kgtkf.tex}
  \caption{Plot of $\IM(G_{--})$ as a function of $\omega$ for $mL=0.3$ for
  extremal $AdS_4$-RN. The top plot is at fixed $k=0.6\,r_h/L^2$ which is less
  than our estimated value of $k_\text{F}\simeq(0.65308\pm0.00009)\,r_h/L^2$ for
  this mass and the bottom plot is for $k=0.7\,r_h/L^2$.}
  \label{Gplot}
\end{figure}

In \cite{lmv} the Fermi momentum for zero fermion mass was determined to
be $k_\text{F}(0)\simeq0.918528499(1)\,r_h/L^2$. In \cite{hh} it was shown that
this value can be found as the solution to an equation involving hypergeometric
functions. We are interested in determining the Fermi momentum for other values
of the mass parameter. In figure \ref{Gplot} we plot $\IM(G_{--})$ for $mL=0.3$
as a function of the frequency for two different values of the momentum. As $k$
approaches $k_\text{F}$ from below, the peak in the top plot in
figure \ref{Gplot} becomes sharper and approaches $\omega=0$. Above $k_\text{F}$
the height of the peak is reduced to two bumps as seen in the bottom plot in
figure \ref{Gplot}. In \cite{lmv} the Fermi momentum was determined by following
the location of the peak to $\omega=0$. This procedure is somewhat laborious,
so we used a numerical routine to automate its determination, at the expense of
reduced precision. We solve the equation of motion \eqref{pnlin} for small but
non-zero $\omega$ and $k<k_\text{F}$. We do this repeatedly for $k<k_\text{F}$,
for each $k$ determining the frequency $\omega^*$ where $\IM(G_{--})$ is
maximum. We then suppose that near $k_\text{F}$
\begin{equation}\label{fit}
  \IM(G_{--}(\omega^*,k))\approx \frac{c}{(k_\text{F}-k)^\alpha} \, ,
\end{equation}
to obtain an estimate of $k_\text{F}$.

\begin{figure}
  %\scalebox{0.6}{\input{log_kscale.tex}}
  \input{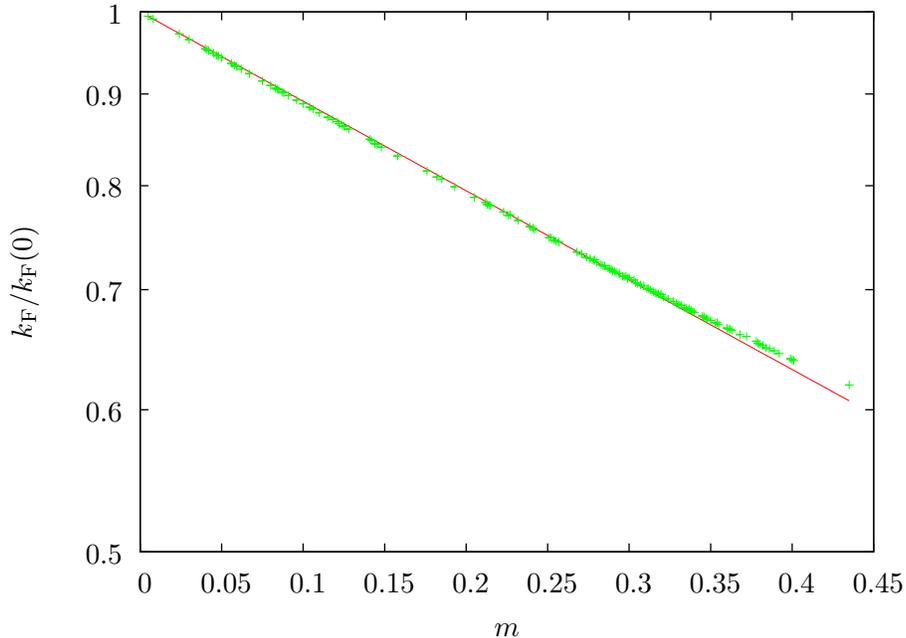}
  \caption{Plot of $k_\text{F}/k_\text{F}(0)$ vs $m$ for extremal $AdS_4$-RN.
  The line is a fit to the data in the form of \eqref{sigma}.}
  \label{kplot}
\end{figure}

In figure \ref{kplot} we plot the ratio of the Fermi momentum at a given mass
to the Fermi momentum at zero mass as determined in \cite{lmv}. Our numerical
routine yielded results that fell into two distinct groupings that differed
by a few orders of magnitude in the goodness of fit to the extrapolation of the
pole \eqref{fit}. We kept only those points in the group with the better fit.
Among this grouping it also yielded two groups of values of $k_\text{F}$ 
that were determined to about $10^{-2}$ and those determined to about $10^{-4}$.
We kept only those values of $k_\text{F}$ that were determined to better than
$10^{-4}$. We have plotted the ratio $k_\text{F}/k_\text{F}(0)$ on a logarithmic
scale to emphasize that the Fermi momentum appears to fall off exponentially
from the value it takes at zero fermion mass. We determined the rate of fall-off
to be
\begin{equation} \label{sigma}
  k_\text{F}\approx k_\text{F}(0)\exp(-\sigma mL), \qquad
  \sigma\simeq1.148\pm0.002 \, .
\end{equation}

The change in the Fermi momentum occurs at fixed chemical potential, and fixed 
charge density in the boundary theory, as they are determined by the background 
gauge field. As such the change in the Fermi momentum cannot be ascribed to a 
change in the density of charge carriers. The fact that this ratio is not unity for all 
values of the mass indicates a violation of Luttinger's theorem. 

The scaling form \eqref{sigma} suggests that a novel scaling law replaces Luttinger's
theorem. The bulk mass is related to the conformal dimension of the boundary field 
as $\Delta = {d\over 2} + |mL|$, so the scaling relation can be recast as 
$k_F\propto e^{-\sigma\Delta}$. This form does not refer to the holographic set-up
and could be valid for non-Fermi liquids more generally. The parameter $\sigma$ 
should measure violations of the Luttinger theorem in more general settings as well, 
such as the semi-holographic Fermi-liquid \cite{semiholoFL}.  

We repeated the analysis for the $AdS_5$-Reissner-Nordstr\"om black hole,
corresponding to $3+1$ dimensional condensed matter systems. We again
found a violation of Luttinger's theorem of the scaling form \eqref{sigma}, now with 
$k_\text{F}(0)\simeq0.8155$ and $\sigma\simeq0.80$.

%\paragraph*{Discussion}
\section{Discussion}
We have shown that a non-Fermi liquid may violate Luttinger's theorem
in a specific manner. It is interesting to consider how this effect could be
detected in real condensed matter systems. One characteristic manifestation
of the Fermi surface is de Haas-van Alphen effect, the oscillation of the 
magnetic susceptibility as a function of the inverse magnetic field with period
\begin{equation}\label{dHvA}
  \delta\left(\frac1B\right)\propto \frac1{A_\text{FS}} \, ,
\end{equation}
where $A_\text{FS}$ is the area of the Fermi surface. A $B$-field is readily incorporated
by a dyonic black hole (considered in \cite{aj,bhms2,dhs}) and much of the 
computation we performed carries over, with the continuous momentum replaced by 
Landau levels. In particular, \eqref{pnlin} remains the equation of motion after
appropriately replacing $k$. We then expect that the 
exponential relation  \eqref{sigma} between Fermi momentum and mass will 
persist in the presence of a magnetic field. 

For zero mass it is known that adjustment of the magnetic field leads to oscillations in 
the semi-classical theory that can be interpreted as the de Haas-van Alphen 
effect  \cite{bhms2,dhs}. In view of our result, we expect the change in the period of 
de Haas-van Alphen oscillations \eqref{dHvA} to be
\begin{equation}\label{dHvB}
  \delta\left(\frac1B\right)\propto e^{2\Delta\sigma} \, ,
\end{equation}
where $\Delta$ is the scaling dimension of the field, and $\sigma$ is the
scaling constant in \eqref{sigma}. The scaling dimension for various currents
can be determined at vanishing $B$-field and then $\sigma$ can be 
measured from \eqref{dHvB}, at least in principle.

One might also hope to uncover the behavior we have found here using
photo-electric scattering techniques \cite{ARPESrev}. These techniques directly
determine the low energy density of states, and are therefore effective at
measuring the Fermi surface. In systems that are well-described by the
holographic non-Fermi liquid we have studied here, we would expect the size of
the Fermi surface to scale exponentially with the conformal dimension of the
operator with the scaling rate given in \eqref{sigma}.

It is an interesting open question whether the behavior we have seen here
occurs in other holographic settings. Of particular interest is
the background considered in \cite{DHK2}. It was shown there that the low
temperature phase of Einstein-Maxwell-Chern-Simons gravity is dominated by a
solution with vanishing entropy, rather than $AdS_5$-Reissner-Nordstr\"om. Since
that solution satisfies the third law it constitutes an
interesting setting for an examination along the lines of what we have done here.

\begin{acknowledgments}
We would like to thank K.\ Schalm and T. Pereg-Barnea for helpful correspondence
and S.-S.\ Lee and D.\ Vegh for helpful discussions.
\end{acknowledgments}

\end{document}